\documentclass[a4paper,twocolumn,prb,showpacs]{revtex4}

\usepackage{graphicx}
\usepackage{amsmath}

\begin{document}

\title{Elimination of the linearization error in \emph{GW} calculations
based on the \\
linearized augmented-plane-wave method}

\author{Christoph Friedrich}

\email{c.friedrich@fz-juelich.de}

\author{Arno Schindlmayr}

\author{Stefan Bl\"ugel}

\affiliation{Institut f\"ur Festk\"orperforschung, Forschungszentrum J\"ulich, 52425
J\"ulich, Germany}

\author{Takao Kotani}

\affiliation{Department of Chemical and Materials Engineering, Arizona State University,
Tempe, Arizona 85287-6006, USA}

\begin{abstract}
This paper investigates the influence of the basis set on the $GW$
self-energy correction in the full-potential linearized augmented-plane-wave
(LAPW) approach and similar linearized all-electron methods. A systematic
improvement is achieved by including local orbitals that are defined
as second and higher energy derivatives of solutions to the radial
scalar-relativistic Dirac equation and thus constitute a natural extension
of the LAPW basis set. Within this approach linearization errors can
be eliminated, and the basis set becomes complete. While the exchange
contribution to the self-energy is little affected by the increased
basis-set flexibility, the correlation contribution benefits from
the better description of the unoccupied states, as do the quasiparticle
energies. The resulting band gaps remain relatively unaffected, however;
for Si we find an increase of 0.03~eV.
\end{abstract}


\maketitle

\section{Introduction}

Electronic excitation energies may be obtained from the solution of
the quasiparticle equation of many-body perturbation theory. This
equation contains a nonlocal and frequency-dependent operator, the
self-energy $\Sigma^{\mathrm{xc}}(\mathbf{r},\mathbf{r}';\epsilon)$,
which, in principle, incorporates all electronic exchange and correlation
effects. As it cannot be treated exactly for real systems, practical
implementations typically use the $GW$ approximation, \cite{Hedin1965}
which has become increasingly popular for electronic-structure calculations
of excited states in recent years and yields band structures in good
quantitative agreement with experimental spectroscopy for a wide range
of materials.

Due to its technical simplicity, the first implementations were based
on the pseudopotential plane-wave approach. In spite of several approximations
in the numerical treatment, which were necessary because of the lack
of computer power in the 1980s, initial results were very promising.
Hybertsen and Louie \cite{Hybertsen1985} as well as Godby \emph{et
al.} \cite{Godby1986} showed that the calculated band gap of Si fell
within a margin of about 0.1~eV from the experimental value. Shortly
afterwards the same authors reported band gaps for several other semiconducting
materials that turned out to be equally accurate.\cite{Hybertsen1986,Godby1987}
After these pioneering studies the $GW$ approximation was applied
to a variety of semiconductors, insulators, and metals with great
success.\cite{Aulbur2000}

So far, most codes still rely on the pseudopotential approximation,
which restricts the range of materials that can be examined. Transition-metal
compounds and oxides, in particular, cannot be treated efficiently
in this approach. Two early all-electron calculations using the $GW$
approximation were done by Hamada \emph{et al.}\cite{Hamada1990}
for Si and by Aryasetiawan \cite{Aryasetiawan1992} for Ni, both within
the linearized augmented-plane-wave (LAPW) method. \emph{}However,
only very recently were further full-potential implementations reported,
based on the LAPW (Refs.~\onlinecite{Ku2002} and \onlinecite{Usuda2002}),
the linearized muffin-tin orbital (LMTO) (Refs.~\onlinecite{Kotani2002}
and \onlinecite{Faleev2004}), the projector-augmented-wave (PAW)
(Refs.~\onlinecite{Arnaud2000} and \onlinecite{Lebegue2003}), and
the Korringa-Kohn-Rostoker \cite{Ernst2005} method together with
applications to a larger variety of systems. 

Compared to earlier pseudopotential results, it was found, however,
that LAPW, LMTO, and PAW calculations appeared to yield systematically
smaller band gaps for semiconductors and, in many cases, a worse agreement
with experiment. \cite{Arnaud2000,Ku2002,Faleev2004,Lebegue2003,Kotani2002}
Ku and Eguiluz \cite{Ku2002} hence argued that, in contrast to all-electron
methods, the pseudopotential calculations benefited from a fortuitious
error cancellation between the pseudopotential approximation and the
neglect of vertex corrections in the $GW$ approximation. Although
their calculations were subsequently criticized for not being converged
with respect to the number of bands,\cite{Delaney2004,Tiago2004}
other all-electron calculations showed a similar underestimation of
the band gap.\cite{Arnaud2000,Kotani2002,Faleev2004,Lebegue2003}
In a different attempt to make one step towards an all-electron treatment,
Tiago \emph{et al.}\cite{Tiago2004} relaxed the pseudopotential approximation
by constructing a pseudopotential only for the 1\emph{s} state of
Si while treating the 2\emph{s} and 2\emph{p} states as valence. Surprisingly,
the resulting band gaps did not deviate substantially from the previous
pseudopotential results, which made the conjecture of Ku and Eguiluz
doubtful. In order to resolve this conflict, it is imperative to carefully
analyze and compare the numerical approximations made in the two approaches. 

Of course, deviations are expected for several reasons. First, the
pseudized wave functions differ from their true counterparts and modify
the matrix elements of the self-energy. Second, the core electrons
are not included in the construction of the nonlocal self-energy if
pseudopotentials are used, which may lead to errors in the core-valence
exchange contribution. On the other hand, the single-particle wave
functions (and the corresponding energies) of both pseudopotential
and linearized all-electron approaches, such as LAPW or LMTO, become
more and more inaccurate at higher energies. The LAPW basis set, on
which we concentrate in the following, is defined by an expansion
around fixed energy parameters that yield accurate wave functions
only in their neighborhood, i.e., the valence band. Of course, this
does not affect calculations within density-functional theory\cite{Hohenberg1964}
(DFT), which only makes use of the occupied states. The $GW$ self-energy,
however, depends on the unoccupied states up to high energies through
the Green function $G$ as well as the screened Coulomb interaction
$W$. The inappropriate description of these states might, therefore,
cause errors in the self-energy correction.

The construction of the pseudopotentials guarantees an accurate wave
function and energy only for the ground state of a given angular momentum
but leads to deviations for higher-lying states. While the deficiency
in the pseudopotential approach is inherent in the pseudopotential
construction, in the LAPW method it must be attributed to the inadequacy
of the basis set for high-lying unoccupied states and can be overcome
by increasing the basis-set flexibility. It is the purpose of this
paper to elucidate the influence of the basis set on the \textbf{$GW$}
results by systematically extending it towards basis-set completeness.
We achieve this by adding local orbitals defined as second and higher
energy derivatives of solutions of the radial scalar-relativistic
Dirac equation. There are several alternative approaches: Bross and
Fehrenbach\cite{Bross1990} use spline functions to augment the basis
set, Krasovskii \emph{et al.}\cite{Krasovskii1994} employ conventional
local orbitals located in the conduction bands together with their
energy derivatives. The advantage of the present approach is that
no special consideration about the energy parameters is needed. Furthermore,
with the order of the derivatives in each angular-momentum channel,
it contains well defined convergence parameters that allow a systematic
attainment of basis-set completeness.

This paper is organized as follows. In Sec.~\ref{sec:Methods} we
provide a brief introduction to the $GW$ and LAPW methods. In Sec.~\ref{sec:FLAPWext}
we describe our extension of the LAPW basis set in detail. As an illustration,
in Sec.~\ref{sec:GW} the effect of the basis-set extension on the
$GW$ results for Si is discussed. We show that the extension of the
basis set and the convergence with respect to the number of bands
both reduce the discrepancy with the pseudopotential and the experimental
band gap. Unless stated otherwise, we use Hartree atomic units.

\section{Methods\label{sec:Methods}}

\subsection{\emph{GW} approximation}

Within many-body perturbation theory, the quasiparticle wave functions
$\psi_{n\mathbf{k}\sigma}(\mathbf{r})$ and energies $\epsilon_{n\mathbf{k}\sigma}$
are obtained from the solution of the quasiparticle equation\begin{eqnarray}
\lefteqn{\left(-\frac{1}{2}\nabla^{2}+V^{\mathrm{ext}}(\mathbf{r})+V^{\mathrm{H}}(\mathbf{r})\right)\psi_{n\mathbf{k}\sigma}(\mathbf{r})}\label{Eq:qp-equ}\\
 & +\int\Sigma_{\sigma}^{\mathrm{xc}}(\mathbf{r},\mathbf{r}';\epsilon_{n\mathbf{k}\sigma})\psi_{n\mathbf{k}\sigma}(\mathbf{r}')\, d^{3}r' & =\epsilon_{n\mathbf{k}\sigma}\psi_{n\mathbf{k}\sigma}(\mathbf{r})\;,\nonumber \end{eqnarray}
where $V^{\mathrm{ext}}(\mathbf{r})$, $V^{\mathrm{H}}(\mathbf{r})$,
and $\Sigma_{\sigma}^{\mathrm{xc}}\left(\mathbf{r},\mathbf{r}';\epsilon_{n\mathbf{k}\sigma}\right)$
are the external potential created by the crystal field and other
applied static fields, the Hartree potential, and the exchange-correlation
self-energy, respectively. The quantum numbers $n$, $\mathbf{k}$,
and $\sigma$ signify the band, wave vector, and spin. Like the majority
of existing implementations, our code exploits the formal similarity
to the Kohn-Sham (KS) equation to obtain approximate energies within
first-order perturbation theory\begin{equation}
\epsilon_{n\mathbf{k}\sigma}\approx\epsilon_{n\mathbf{k}\sigma}^{\mathrm{KS}}+Z_{n\mathbf{k}\sigma}\left\langle \varphi_{n\mathbf{k}\sigma}\left|\Sigma_{\sigma}^{\mathrm{xc}}\left(\epsilon_{n\mathbf{k}\sigma}^{\mathrm{KS}}\right)-V_{\sigma}^{\mathrm{xc}}\right|\varphi_{n\mathbf{k}\sigma}\right\rangle \;,\end{equation}
where $V_{\sigma}^{\mathrm{xc}}(\mathbf{r})$ is the local exchange-correlation
potential, $\varphi_{n\mathbf{k}\sigma}(\mathbf{r})$ the Kohn-Sham
wave function, and the quasiparticle renormalization factor is given
by\begin{equation}
Z_{n\mathbf{k}\sigma}=\left(1-\left\langle \varphi_{n\mathbf{k}\sigma}\left|\frac{\partial\Sigma_{\sigma}^{\mathrm{xc}}}{\partial\epsilon}\left(\epsilon_{n\mathbf{k}\sigma}^{\mathrm{KS}}\right)\right|\varphi_{n\mathbf{k}\sigma}\right\rangle \right)^{-1}\le1\,.\label{eq:renormalization}\end{equation}
 In the following we will always use the shorthand notation $\left\langle \Sigma^{\mathrm{xc}}\right\rangle =\left\langle \varphi_{n\mathbf{k}\sigma}\left|\Sigma_{\sigma}^{\mathrm{xc}}(\epsilon_{n\mathbf{k}\sigma}^{\mathrm{KS}})\right|\varphi_{n\mathbf{k}\sigma}\right\rangle $.

We employ the $GW$ approximation for the self-energy, symbolically
written as $\Sigma^{\mathrm{xc}}=iGW$, where $G$ is the Kohn-Sham
Green function\begin{eqnarray}
G_{\sigma}(\mathbf{r},\mathbf{r}';\epsilon) & = & \sum_{n,\mathbf{k}}^{\mathrm{occ}}\frac{\varphi_{n\mathbf{k}\sigma}(\mathbf{r})\varphi_{n\mathbf{k}\sigma}^{*}(\mathbf{r}')}{\epsilon-\epsilon_{n\mathbf{k}\sigma}-i\delta}\nonumber \\
 &  & +\sum_{n,\mathbf{k}}^{\mathrm{unocc}}\frac{\varphi_{n\mathbf{k}\sigma}(\mathbf{r})\varphi_{n\mathbf{k}\sigma}^{*}(\mathbf{r}')}{\epsilon-\epsilon_{n\mathbf{k}\sigma}+i\delta}\label{eq:Green}\end{eqnarray}
($\delta$ is an infinitesimal positive number) and $W$ the dynamically
screened Coulomb interaction. The latter is calculated from $W=v+vPW$,
where $v$ is the bare Coulomb potential and $P$ the polarization
function\begin{eqnarray}
P_{\sigma}(\mathbf{r},\mathbf{r}';\epsilon) & = & \sum_{n,\mathbf{k}}^{\mathrm{occ}}\sum_{n',\mathbf{k}'}^{\mathrm{unocc}}\frac{2\left(\epsilon_{n'\mathbf{k}'\sigma}-\epsilon_{n\mathbf{k}\sigma}-i\delta\right)}{\epsilon^{2}-\left(\epsilon_{n'\mathbf{k}'\sigma}-\epsilon_{n\mathbf{k}\sigma}-i\delta\right)^{2}}\label{Eq:polarization}\\
 &  & \times\varphi_{n\mathbf{k}\sigma}^{*}(\mathbf{r})\varphi_{n'\mathbf{k}'\sigma}(\mathbf{r})\varphi_{n'\mathbf{k}'\sigma}^{*}(\mathbf{r}')\varphi_{n\mathbf{k}\sigma}(\mathbf{r}')\nonumber \end{eqnarray}
in the random-phase approximation. In practice, the self-energy is
decomposed into exchange and correlation contributions \begin{equation}
\Sigma^{\mathrm{xc}}=\Sigma^{\mathrm{x}}+\Sigma^{\mathrm{c}}=iGv+iG(W-v)\;.\label{eq:decomposition}\end{equation}
The exchange contribution only depends on occupied states, whereas
unoccupied states up to high energies, typically $100-200$~eV above
the Fermi energy, are needed for an accurate evaluation of the correlation
contribution.

\subsection{LAPW basis set}

The all-electron APW method\cite{Slater1937} as well as the related
LAPW method\cite{Andersen1975} rely on a decomposition of space into
muffin-tin (MT) spheres, centered at the atomic nuclei, and the interstitial
region. The core-electron wave functions, which are (mostly) confined
to the muffin-tin spheres, are directly obtained from a solution of
the fully relativistic Dirac equation. Here only the spherical part
of the effective potential is retained. For the valence electrons
a basis set $\left\{ \phi_{\mathbf{k}+\mathbf{G}}^{\sigma}(\mathbf{r})\right\} $
is constructed. Its basis functions, the so-called augmented plane
waves, are defined everywhere in space. The smoothness of the potential
in the interstitial region motivates the use of plane waves with $|\mathbf{k}+\mathbf{G}|\leq K_{\mathrm{max}}$,
where $K_{\mathrm{max}}$ is a convergence parameter. In the MT spheres,
on the other hand, the potential is peaked at the nuclei and predominantly
spherical. In the APW method, which is usually implemented in combination
with a shape approximation of spherically symmetric potentials inside
the MT spheres, one uses numerical solutions $u_{l\sigma}^{(0)}(\epsilon_{l\sigma},r)$
of the radial Schr\"odinger equation\begin{equation}
\hat{h}_{l\sigma}ru_{l\sigma}^{(0)}(\epsilon_{l\sigma},r)=\epsilon_{l\sigma}ru_{l\sigma}^{(0)}(\epsilon_{l\sigma},r)\label{eq:u-gen}\end{equation}
with the Kohn-Sham Hamiltonian\begin{equation}
\hat{h}_{l\sigma}=-\frac{1}{2}\frac{\partial^{2}}{\partial r^{2}}+\frac{l(l+1)}{2r^{2}}+V_{\sigma}^{\mathrm{eff}}(r)\end{equation}
 and the relevant effective potential $V_{\sigma}^{\mathrm{eff}}(r)$.
For simplicity we give the nonrelativistic equations here; the scalar-relativistic
versions are shown in the Appendix. The atom index is suppressed throughout.
Augmenting the interstitial plane waves with linear combinations of
the $u_{lm\sigma}^{(0)}(\epsilon_{l\sigma},\mathbf{r})=u_{l\sigma}^{(0)}(\epsilon_{l\sigma},r)Y_{lm}(\hat{\mathbf{r}})$,
taking into account the continuity at the MT sphere boundaries, yields
the basis functions $\phi_{\mathbf{k}+\mathbf{G}}^{\sigma}(\mathbf{r})$.
In the APW method the energy parameters $\epsilon_{l\sigma}$ are
identical to the band energies $\epsilon_{n\mathbf{k}\sigma}$. For
a spherically symmetric MT potential this condition guarantees that
the Kohn-Sham wave functions $\varphi_{n\mathbf{k}\sigma}(\mathbf{r})$
can be obtained exactly (subject to convergence with the \textbf{$l$}
cutoff and $K_{\mathrm{max}}$), but it leads to an expensive nonlinear
eigenvalue problem. In the LAPW method this difficulty is circumvented
by fixing the $\epsilon_{l\sigma}$ at suitable energies in the valence-band
region, thereby avoiding the expensive self-consistency condition
with respect to the band energies and making a linear solution of
the Kohn-Sham eigenvalue equation possible. Apart from solutions $u_{l\sigma}^{(0)}(\epsilon_{l\sigma},r)$
of the radial Schr\"odinger equation (\ref{eq:u-gen}), the energy derivatives
$u_{l\sigma}^{(1)}(\epsilon_{l\sigma},r)=\partial u_{l\sigma}^{(0)}(\epsilon_{l\sigma},r)/\partial\epsilon$
are also employed to increase the basis-set flexibility in the MT
spheres. They are obtained from \begin{subequations}\label{eqall:udot-gen}\begin{equation}
\hat{h}_{l\sigma}ru_{l\sigma}^{(1)}(\epsilon_{l\sigma},r)=\epsilon_{l\sigma}ru_{l\sigma}^{(1)}(\epsilon_{l\sigma},r)+ru_{l\sigma}^{(0)}(\epsilon_{l\sigma},r)\;,\end{equation}
\begin{equation}
\frac{d}{d\epsilon}\langle u_{l\sigma}^{(0)}(\epsilon_{l\sigma})|u_{l\sigma}^{(0)}(\epsilon_{l\sigma})\rangle=2\langle u_{l\sigma}^{(0)}(\epsilon_{l\sigma})|u_{l\sigma}^{(1)}(\epsilon_{l\sigma})\rangle=0\;.\label{eq:udot-gen-condition}\end{equation}
\end{subequations}The two sets of radial functions are combined with
the interstitial plane waves in such a way that not only the basis
functions but also their radial derivatives are continuous at the
MT sphere boundaries. This procedure yields the LAPW basis set for
the valence electrons \begin{equation}
\phi_{\mathbf{k}+\mathbf{G}}^{\sigma}(\mathbf{r})=\left\{ \begin{array}{ll}
{\displaystyle \frac{1}{\sqrt{\Omega}}}e^{i(\mathbf{k}+\mathbf{G})\cdot\mathbf{r}} & \textrm{if }\mathbf{r}\notin\mathrm{MT},\\
{\displaystyle \sum_{l,m}\sum_{\nu=0}^{1}}a_{lm\sigma\nu}^{\mathbf{k}+\mathbf{G}}u_{lm\sigma}^{(\nu)}(\epsilon_{l\sigma},\mathbf{r}) & \textrm{if }\mathbf{r}\in\mathrm{MT},\end{array}\right.\label{Eq:FLAPW-basis}\end{equation}
where the coefficients $a_{lm\sigma\nu}^{\mathbf{k+G}}$ are uniquely
determined by the matching conditions and $\Omega$ is the unit-cell
volume. The inclusion of the energy derivatives implies a linear approximation
for the radial functions\begin{equation}
u_{l\sigma}\left(\epsilon_{n\mathbf{k}\sigma},r\right)\approx u_{l\sigma}^{(0)}\left(\epsilon_{l\sigma},r\right)+\left(\epsilon_{n\mathbf{k}\sigma}-\epsilon_{l\sigma}\right)u_{l\sigma}^{(1)}\left(\epsilon_{l\sigma},r\right)\,,\label{eq:Taylor}\end{equation}
thus introducing a linearization error, which grows with the deviation
of the band energy $\epsilon_{n\mathbf{k}\sigma}$ from the parameters
$\epsilon_{l\sigma}$ and becomes especially relevant for high-lying
states. In full-potential LAPW implementations the radial functions
are still derived from the spherical part of the MT potential. Therefore,
the actual wave functions cannot be constructed from the $u_{lm\sigma}^{(0)}(\epsilon_{l\sigma},\mathbf{r})$
alone, but as the nonspherical modulation is typically small, it is
commonly accepted that the energy derivatives add enough basis-set
flexibility to describe the valence electrons in the full potential
accurately.\cite{Wimmer1981}

In order to quantify the error incurred by the approximation (\ref{eq:Taylor}),
we compare the Kohn-Sham eigenvalue spectra with and without the linearization
of the radial functions. In the second case we determine the band
energies $\epsilon_{n\mathbf{k}\sigma}$ iteratively by setting the
energy parameters $\epsilon_{l\sigma}$ equal to the $\epsilon_{n\mathbf{k}\sigma}$
until self-consistency is reached. For a spherically symmetric MT
potential this procedure is equivalent to the APW method, because
the contribution of the energy derivatives in our scheme vanishes
at the self-consistency point $\epsilon_{l\sigma}=\epsilon_{n\mathbf{k}\sigma}$,
and the description of the wave functions is thus identical: linear
combinations of $u_{lm\sigma}^{(0)}(\epsilon_{l\sigma},\mathbf{r})$
inside the MT spheres and plane waves in the interstitial region.
It should be noted that although the APW basis functions exhibit a
derivative discontinuity at the MT sphere boundaries, the wave functions
themselves are smooth. Furthermore, the increased basis-set flexibility
achieved by the additional radial functions makes it possible to apply
our method to the full crystal potential. In this case the energy
derivatives yield a nonvanishing but small contribution that reflects
the nonspherical modulation. 

\begin{figure}
\includegraphics[%
  clip,
  width=0.82\columnwidth,
  angle=-90]{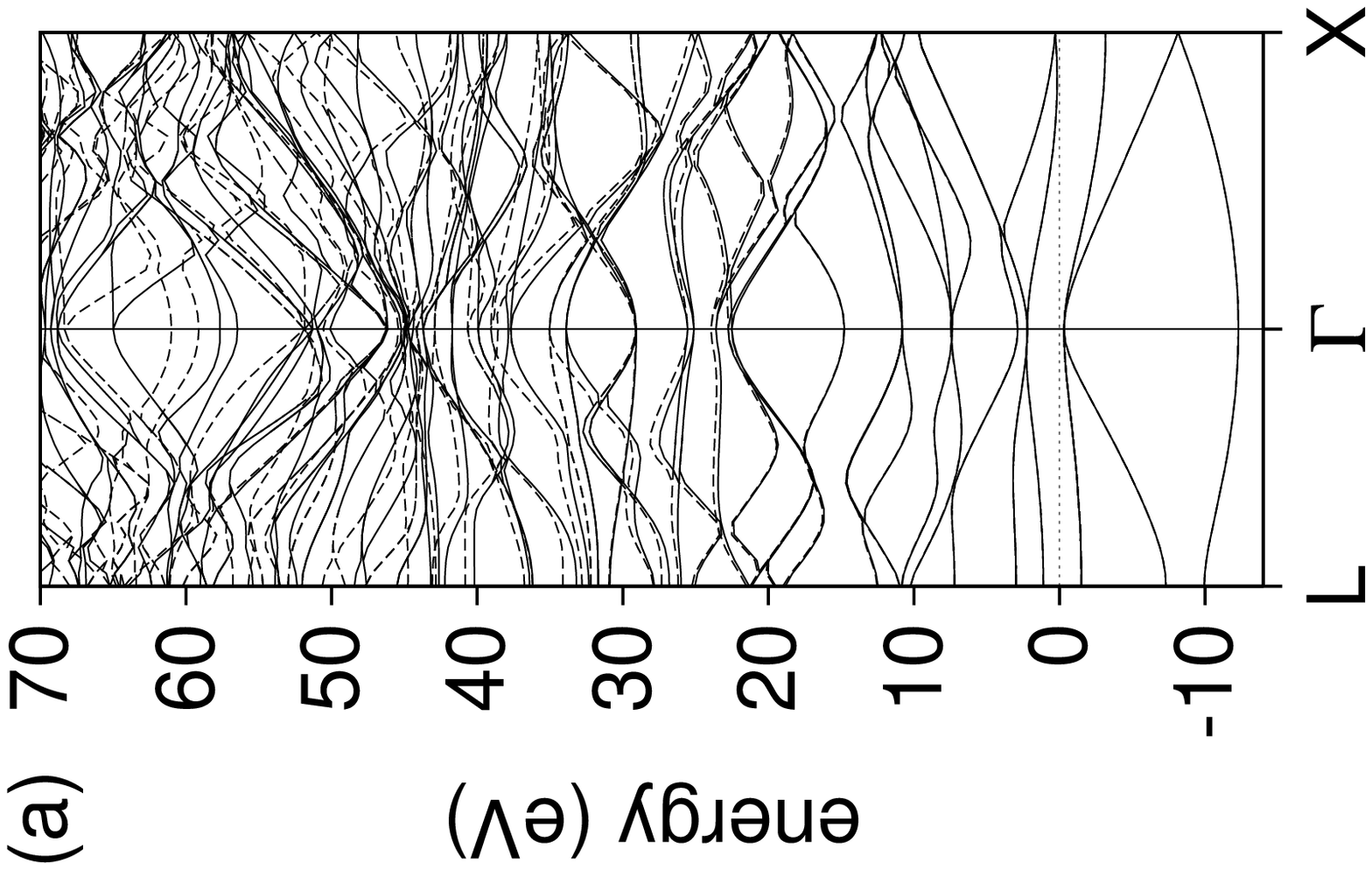}\includegraphics[%
  clip,
  width=0.82\columnwidth,
  angle=-90]{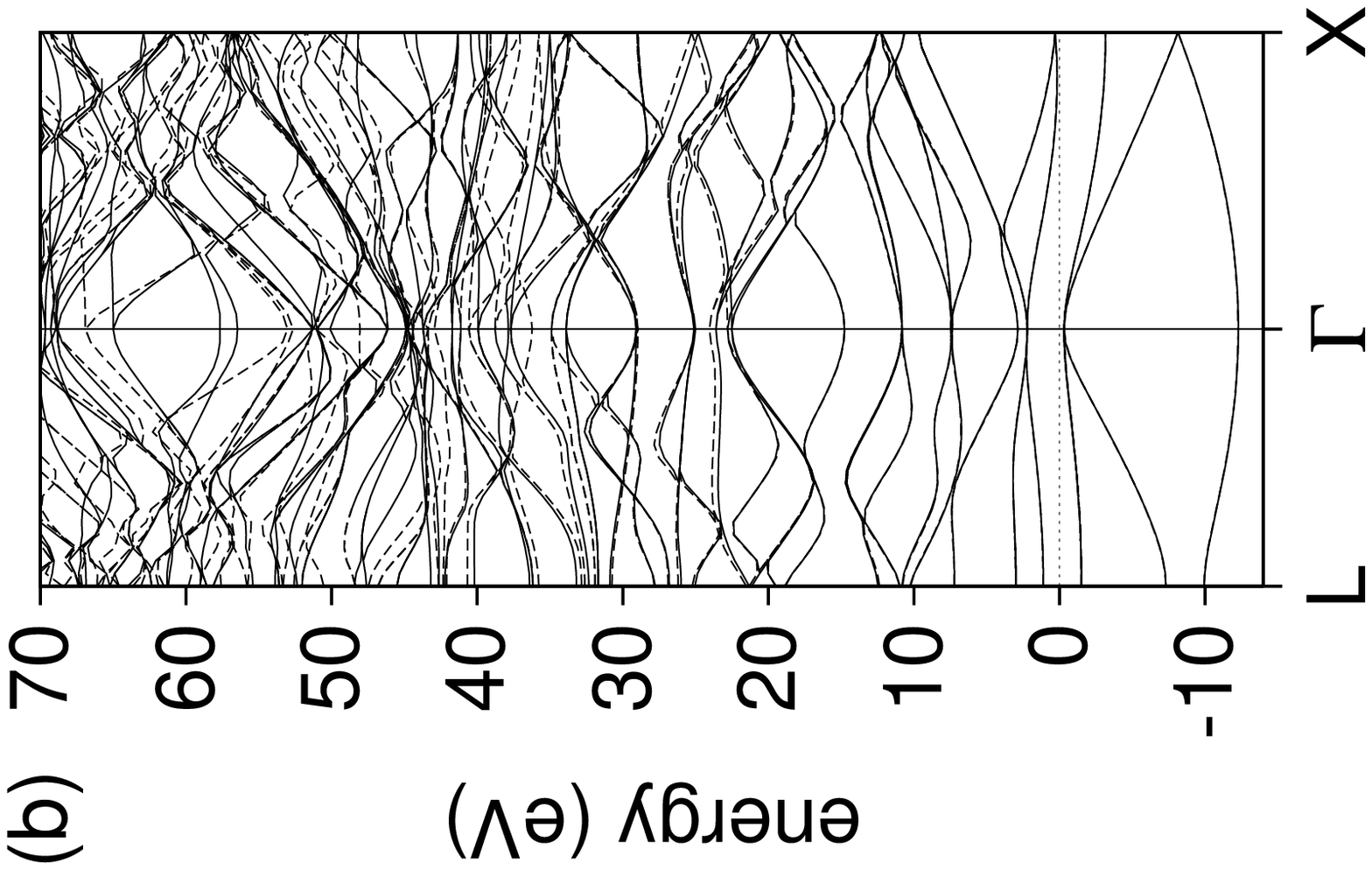}

\caption{\label{cap:bandstruc}Comparison of (a) full-potential LAPW and (b)
pseudopotential band structures for silicon in the local-density approximation
(dashed lines) with results from our nonlinear scheme in which the
energy parameters are determined self-consistently (solid lines).
In both cases deviations start to appear at around 20~eV above the
Fermi energy (0~eV).}
\end{figure}

In Fig.~\ref{cap:bandstruc}(a) we compare the resulting Kohn-Sham
band structure with that from a standard full-potential LAPW calculation
for the prototype semiconductor Si. %
 Both band structures are evaluated within the local-density approximation
for the same effective potential, which is obtained from the LAPW
self-consistency loop. The calculation is carried out with the experimental
lattice constant of 10.26~Bohr, a muffin-tin radius of 2.16~Bohr,
$K_{\mathrm{max}}=4.0$~Ha, an angular-momentum cutoff of 8, and
512 $\mathbf{k}$ points in the full Brillouin zone. We use the \texttt{FLEUR}
code. \cite{Fleur} In Fig.~\ref{cap:bandstruc}(b) we make a similar
comparison with the pseudopotential plane-wave method, which is the
most frequently used approach in the context of the $GW$ approximation.
For this calculation we use a standard norm-conserving Hamann pseudopotential
\cite{Hamann1989} and an energy cutoff of 9~Ha for the plane waves,
the other parameters are the same as in LAPW\@. In both cases deviations
start to occur at around 20~eV above the Fermi energy, i.e., around
the twentieth band. This is far below the number of bands normally
included in a $GW$ calculation to ensure convergence of the self-energy.

In the next section we discuss our extension of the LAPW basis set,
which allows a systematic reduction and ultimate elimination of the
linearization error. In the basis-set limit the basis becomes complete
and all states, including high-lying conduction bands, are accurately
described.

\section{The LAPW basis-set extension\label{sec:FLAPWext}}

In order to enhance the basis-set flexibility, it is not sufficient
to simply increase the plane-wave cutoff $K_{\mathrm{max}}$, as is
the case in pseudopotential calculations, since this only improves
the basis set in the interstitial region but not within the muffin-tin
spheres. In fact, Krasovskii\cite{Krasowski1997} has shown that a
fairly small $K_{\mathrm{max}}$ is sufficient to provide enough flexibility
in the interstitial region, while the MT part of the basis set quickly
deteriorates the more the wave-function energies deviate from the
parameters $\epsilon_{l\sigma}$. This inadequacy cannot be overcome
with a higher $K_{\mathrm{max}}$.

In a straightforward extension of the LAPW approach one can include
second energy derivatives in the Taylor expansion (\ref{eq:Taylor}).
In this case one would need a third matching condition, though, in
order to uniquely define the augmented plane waves (\ref{Eq:FLAPW-basis}),
e.g., the continuity of the second radial derivative at the MT sphere
boundaries. These more stringent conditions are known to lead to a
less efficient basis set, i.e., one with a slower convergence with
respect to $K_{\mathrm{max}}$, however. \cite{Singh1991} To circumvent
this problem we introduce the higher energy derivatives as additional
basis functions in the form of local orbitals, \cite{Singh1991} which
vanish outside and on the MT sphere boundaries and, therefore, do
not need to be matched to plane waves. In general, the radial part
of a local orbital is constructed as a linear combination of three
radial functions with the conditions of vanishing value and derivative
at the MT sphere boundary as well as normalization. In our approach
these three functions are $u_{l\sigma}^{(0)}(\epsilon_{l\sigma},r)$,
$u_{l\sigma}^{(1)}(\epsilon_{l\sigma},r),$ and one of $u_{l}^{(\nu)}(\epsilon_{l\sigma},r)$
with $\nu\ge2$. The latter is the $\nu$th energy derivative, which
can be obtained in analogy to Eq.~(\ref{eqall:udot-gen}) from \begin{subequations}\label{eqall:ulo-gen}\begin{equation}
\hat{h}_{l\sigma}ru_{l\sigma}^{(\nu)}(\epsilon_{l\sigma},r)=\epsilon_{l\sigma}ru_{l\sigma}^{(\nu)}(\epsilon_{l\sigma},r)+\nu ru_{l\sigma}^{(\nu-1)}(\epsilon_{l\sigma},r)\;,\end{equation}
\begin{equation}
\frac{d^{\nu}}{d\epsilon^{\nu}}\langle u_{l\sigma}^{(0)}(\epsilon_{l\sigma})|u_{l\sigma}^{(0)}(\epsilon_{l\sigma})\rangle=0\end{equation}
\end{subequations}(see the Appendix for the scalar-relativistic treatment).

We now prove that the functions $u^{(\nu)}(\epsilon,r)$ are (i) linearly
independent of each other and (ii) orthogonal to solutions $u^{(0)}(\epsilon_{0},r)$
of Eq.~(\ref{eq:u-gen}) with $\epsilon_{0}\neq\epsilon$ that vanish
outside and on the MT sphere boundary. For simplicity the indices
$l$ and $\sigma$ are omitted here.

The first statement guarantees that each derivative increases the
variational freedom and does not lead to an overcomplete basis set.
For $u^{(0)}(\epsilon,r)$ and $u^{(1)}(\epsilon,r)$ this follows
trivially from condition (\ref{eq:udot-gen-condition}). Now assume
that $u^{(\nu)}(\epsilon,r)$ for some $\nu\ge2$ can be written as
a linear combination\begin{equation}
u^{(\nu)}(\epsilon,r)=\sum_{\mu=0}^{\nu-1}c_{\mu}u^{(\mu)}(\epsilon,r)\end{equation}
 of the linearly independent functions $u^{(\mu)}(\epsilon,r)$ with
$0\le\mu\le\nu-1$. Then the application of Eqs.~(\ref{eq:u-gen}),
(\ref{eqall:udot-gen}), and (\ref{eqall:ulo-gen}) leads to \begin{eqnarray}
\nu ru^{(\nu-1)}(\epsilon,r) & = & \left(\hat{h}-\epsilon\right)ru^{(\nu)}(\epsilon,r)\nonumber \\
 & = & \sum_{\mu=0}^{\nu-2}\mu c_{\mu+1}ru^{(\mu)}(\epsilon,r)\;,\end{eqnarray}
 which contradicts the assumption of linear independence of the $u^{(\mu)}(\epsilon,r)$
with $0\le\mu\le\nu-1$. Therefore, the derivatives of orders $0,\ldots,\nu$
must all be linearly independent.

The second statement guarantees that orbitals $u^{(\nu)}(\epsilon,r)$
constructed from the valence basis are orthogonal to the core states
$u^{(0)}(\epsilon_{0},r)$. It should be noted, however, that shallow
semicore states do not vanish sufficiently outside the MT spheres
and should, therefore, be treated with (conventional) local orbitals
at suitable energies. For the special case $\nu=0$ the second statement
follows from $(\hat{h}-\epsilon_{0})ru^{(0)}(\epsilon_{0},r)=(\hat{h}-\epsilon)ru^{(0)}(\epsilon,r)=0$
with $\epsilon\ne\epsilon_{0}$ and integration by parts, giving \begin{eqnarray}
\lefteqn{(\epsilon-\epsilon_{0})\left\langle u^{(0)}(\epsilon_{0})\right|\left.u^{(0)}(\epsilon)\right\rangle }\nonumber \\
 & = & \int_{0}^{R}\left[ru^{(0)}(\epsilon_{0},r)\left(\hat{h}ru^{(0)}(\epsilon,r)\right)\right.\nonumber \\
 &  & \left.-\left(\hat{h}ru^{(0)}(\epsilon_{0},r)\right)ru^{(0)}(\epsilon,r)\right]dr\nonumber \\
 & = & \frac{R^{2}}{2}\left(u^{(0)}(\epsilon,R)u^{(0)\prime}(\epsilon_{0},R)-u^{(0)}(\epsilon_{0},R)u^{(0)\prime}(\epsilon,R)\right)\nonumber \\
 & = & 0\;,\end{eqnarray}
where $R$ is the MT radius. For $\nu\ge1$ this expression instead
reads \begin{eqnarray}
\lefteqn{(\epsilon-\epsilon_{0})\left\langle u^{(0)}(\epsilon_{0})\right|\left.u^{(\nu)}(\epsilon)\right\rangle }\nonumber \\
 & = & \frac{R^{2}}{2}\left(u^{(\nu)}(\epsilon,R)u^{(0)\prime}(\epsilon_{0},R)-u^{(0)}(\epsilon_{0},R)u^{(\nu)\prime}(\epsilon,R)\right)\nonumber \\
 &  & -\nu\left\langle u^{(0)}(\epsilon_{0})\right|\left.u^{(\nu-1)}(\epsilon)\right\rangle =0\;,\end{eqnarray}
where the last equality follows from induction. 

\begin{figure}
\includegraphics[%
  width=0.70\columnwidth,
  angle=-90]{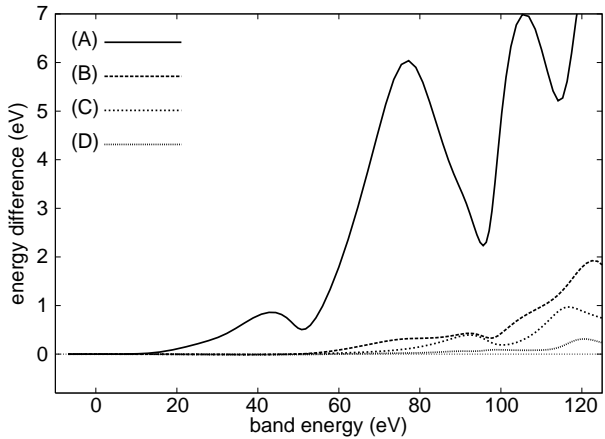}

\caption{\label{cap:FLAPWLO}Deviation of the Kohn-Sham eigenvalues from the
results of our nonlinear reference calculation at the $\Gamma$ point
of Si as a function of energy for the basis sets (A), (B), (C), and
(D) (for details see the text). The energy parameters are optimized
for the valence bands and identical in all calculations. The eigenvalues
are given with respect to the Fermi level. The curves are smoothed
using a Bézier algorithm for clarity.}
\end{figure}
When applied to Si, the example shown in Fig.~\ref{cap:bandstruc},
already the inclusion of second-derivative local $s$, $p$, $d$,
and $f$ orbitals yields agreement with the reference band structure
from our nonlinear scheme, on the scale of the figure, up to 60~eV
above the Fermi energy. The addition of third derivatives and \emph{g}
functions pushes this limit up to around 80~eV. The improvement over
conventional LAPW in the description of the unoccupied states is clearly
seen in Fig.~\ref{cap:FLAPWLO}, which shows the deviation of the
single-particle energies from the corresponding reference values for
(A) the conventional LAPW basis set, (B) with second-derivative local
orbitals for $l\le3$, (C) with second-derivative local orbitals for
$l\le4$, and (D) with second- and third-derivative local orbitals
for $l\le4$ at the $\Gamma$ point of Si as a function of the single-particle
energies. The inclusion of just the second-derivative local orbitals
for all $l\leq3$ in (B) gives energies within 0.2~eV of the reference
values up to 60~eV above the Fermi energy. Adding local orbitals
with $l=4$ in (C) and third derivatives in (D) further improves the
agreement. The advantage over similar approaches to improve the flexibility
of the basis set (e.g., Ref.~\onlinecite{Krasovskii1994}) is that
systematic convergence can be achieved without special assumptions
about the energy parameters of the additional local orbitals. As we
find the results to be relatively independent of their position, one
can simply use the LAPW energy parameters, which are located at the
center of gravity of the valence band. The convergence towards basis-set
completeness is then controlled by two simple parameters, the maximum
quantum number $l$ of the local orbitals and the order of derivatives
$\nu$ (together with the $l$ cutoff and $K_{\mathrm{max}}$ of the
augmented plane waves).

\section{Application to $\textrm{\boldmath$GW$}$\label{sec:GW}}

Once the wave functions are determined in the basis (\ref{Eq:FLAPW-basis}),
the polarization function (\ref{Eq:polarization}) and related quantities
are represented in terms of a mixed basis designed for the expansion
of products of eigenfunctions.\cite{Kotani2002} It consists of plane
waves with a cutoff of 2.7~$\textrm{Bohr}^{-1}$ in the interstitial
region and products of two radial basis functions inside the MT spheres,
where the first is related to an occupied and the second to an unoccupied
state. For occupied states $u_{l}^{(1)}$ contributions and higher
energy derivatives can be neglected, since the parameter $\epsilon_{l}$
is chosen to be the center of gravity of the occupied $l$-like density
of states ($\epsilon_{0}=-7.9$~eV, $\epsilon_{1}=-3.2$~eV, $\epsilon_{2}=-3.4$~eV,
$\epsilon_{l\ge3}=-4.8$~eV relative to the top of the valence band).
For unoccupied states the inclusion of $u_{l}^{(\nu)}$ with $\nu\ge1$
as well as higher angular momenta can be important, because the corresponding
wave-function coefficients become large for higher-lying states. In
our calculations the mixed basis is constructed from $u_{l}^{(0)}$
with $l\le2$ for occupied as well as $u_{l}^{(0)}$ with $0\le l\le6$
and $u_{l}^{(\mu)}$ with $1\le\mu\le\nu$ and $l\le L$ {[}(A) $\nu=1,$
$L=3;$ (B) $\nu=2$, $L=3$; (C) $\nu=2$, $L=4$; (D) $\nu=3$,
$L=4${]} for unoccupied states. The omission of the other radial
functions results in a negligible error not exceeding 0.1~meV in
the correlation contribution $\left\langle \Sigma^{\mathrm{c}}\right\rangle $
for all states considered here. The self-energy is evaluated with
an angular-momentum cutoff of 6 and 216 $\mathbf{k}$ points in the
full Brillouin zone. All other parameters are the same as in the underlying
DFT calculation.%
\begin{figure}
\includegraphics[%
  width=1.0\columnwidth,
  angle=-90]{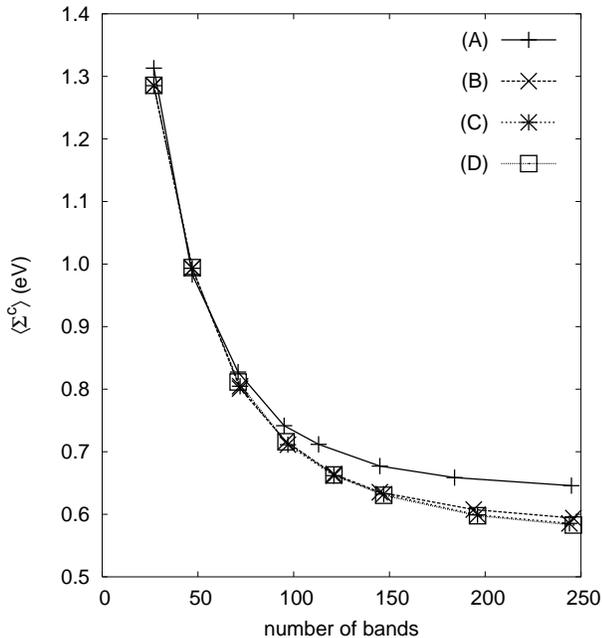}

\caption{\label{cap:Sigmac-curves}Expectation value of the correlation contribution
to the self-energy at the valence-band maximum of Si at $\Gamma$
as a function of the number of bands. The inclusion of second-derivative
local orbitals for $l\le3$ (B) changes the asymptotic value by more
than 0.05~eV with respect to the conventional LAPW basis set (A).
Further extensions of the basis give only minor corrections.}
\end{figure}

Changes in the expectation values of the exchange term $\left\langle \Sigma^{\mathrm{x}}\right\rangle $
and the exchange-correlation potential $\left\langle V^{\mathrm{xc}}\right\rangle $,
which are both independent of the unoccupied states, are small and
compensate each other; the main effect on the quasiparticle energies
is due to the basis-set dependence of the correlation contribution
to the self-energy. In Fig.~\ref{cap:Sigmac-curves} we show $\left\langle \Sigma^{\mathrm{c}}\right\rangle $
for the valence-band maximum of Si at $\Gamma$ as a function of the
number of bands included in the Green function (\ref{eq:Green}) and
the polarization function (\ref{Eq:polarization}) for the different
basis sets (see Sec.~\ref{sec:FLAPWext}). Up to the fiftieth band
the curves are nearly identical but then begin to deviate due to the
linearization error. As the denominator in (\ref{Eq:polarization})
reduces the weight of higher-lying states while their linearization
error grows at the same time, the largest increase of the difference
between the curves is seen for intermediate numbers of bands between
60 and 150. The curves converge rapidly with respect to the basis-set
size: already with the inclusion of second-derivative local orbitals
(B) convergence is reached to within 0.01~eV.

\begin{figure}
\includegraphics[%
  width=1.0\columnwidth,
  angle=-90]{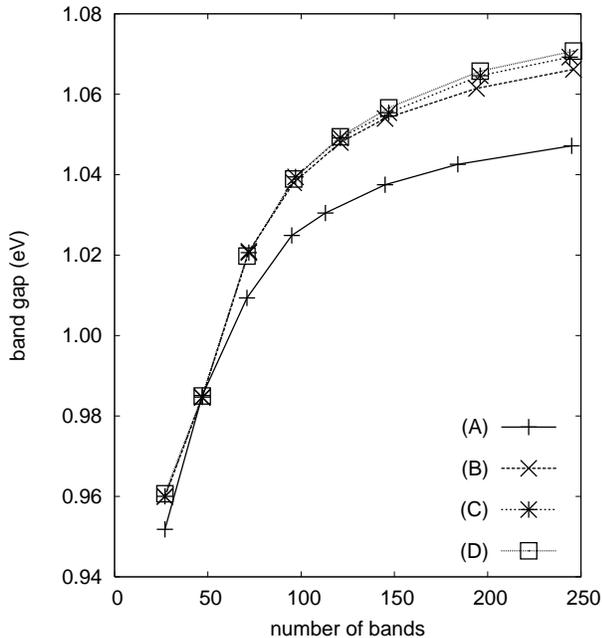}

\caption{\label{cap:bandgap-curves}Indirect band gap of Si as a function
of the number of bands. The inclusion of second-derivative local orbitals
for $l\le3$ (B) provides the largest step towards convergence with
respect to the basis set.}
\end{figure}

\begin{table}

\caption{\label{cap:Sigmac}Expectation values of the correlation contribution
$\left\langle \Sigma^{\mathrm{c}}\right\rangle $ to the self-energy
correction for the basis sets (A), (B), (C), and (D) (see text) together
with the differences $\Delta\left\langle \Sigma^{\mathrm{c}}\right\rangle $
between basis sets (A) and (D). All values are in eV.}

\begin{ruledtabular}

\begin{tabular}{lrrrrr}
&
\multicolumn{4}{c}{$\left\langle \Sigma^{\mathrm{c}}\right\rangle $}&
\tabularnewline
&
\multicolumn{1}{r}{(A)}&
(B)&
(C)&
(D)&
$\Delta\left\langle \Sigma^{\mathrm{c}}\right\rangle $\tabularnewline
\hline
$\Gamma_{25'v}$&
0.646&
0.594&
0.585&
0.583&
--0.063\tabularnewline
$\Gamma_{15c}$&
--4.183&
--4.224&
--4.230&
--4.233&
--0.050\tabularnewline
$X_{4v}$&
1.824&
1.784&
1.778&
  1.776&
--0.049\tabularnewline
$X_{1c}$&
--3.749&
--3.776&
--3.780&
--3.782&
--0.033\tabularnewline
$L_{3'v}$&
  1.139&
1.091&
1.083&
  1.081&
--0.058\tabularnewline
$L_{1c}$&
--3.911&
--3.960&
--3.967&
--3.971&
--0.060\tabularnewline
\end{tabular}

\end{ruledtabular}
\end{table}
\begin{table}

\caption{\label{cap:Quasip}Quasiparticle energies $\epsilon$ for the basis
sets (A), (B), (C), and (D) (see text) together with the differences
$\Delta\epsilon$ between basis sets (A) and (D). All values are in
eV.}

\begin{ruledtabular}

\begin{tabular}{lrrrrr}
&
\multicolumn{4}{c}{$\epsilon$}&
\tabularnewline
&
(A)&
(B)&
(C)&
(D)&
$\Delta\epsilon$\tabularnewline
\hline
$\Gamma_{25'v}$&
--0.646&
--0.685&
--0.692&
--0.694&
--0.048\tabularnewline
$\Gamma_{15c}$&
2.541&
2.514&
2.509&
2.509&
--0.032\tabularnewline
$X_{4v}$&
--3.579&
--3.608&
--3.613&
--3.615&
--0.036\tabularnewline
$X_{1c}$&
0.521&
0.502&
0.499&
0.498&
--0.023\tabularnewline
$L_{3'v}$&
--1.883&
--1.919&
--1.925&
--1.927&
--0.044\tabularnewline
$L_{1c}$&
1.467&
1.432&
1.427&
1.424&
--0.043\tabularnewline
\end{tabular}

\end{ruledtabular}
\end{table}
 The corresponding curves for other states look qualitatively similar.
Tables~\ref{cap:Sigmac} and \ref{cap:Quasip} give the values of
$\left\langle \Sigma^{\mathrm{c}}\right\rangle $ and the resulting
quasiparticle energies $\epsilon$ calculated with 245 bands for the
different basis sets. As a reference, the valence-band maximum in
the underlying Kohn-Sham calculation is set to zero. All quasiparticle
energies tend towards smaller values in the basis-set limit. Consequently,
the effect on relative transition energies is smaller but of the same
order. In Fig.~\ref{cap:bandgap-curves} we show the behavior of
the indirect band gap as an example. %
The results are lowered by 0.02~eV if the $\mathbf{k}$-point mesh
is fully converged, and by another 0.02~eV if screening due to the
2\emph{p} electrons is included in the correlation self-energy. The
convergence with respect to the number of bands and the basis-set
extension both increase the calculated band gap and thus narrow the
distance to the experimental value of 1.17~eV. The final deviation
is comparable to that of typical pseudopotential calculations, which
tend to a slight overestimation.\cite{Hybertsen1985,Godby1986} Although
a certain discrepancy with respect to the pseudopotential results
still remains, we conclude that it is, in fact, smaller than suggested
by previous calculations. \cite{Ku2002,Lebegue2003,Kotani2002,Arnaud2000}
The stronger underestimation in these studies must be attributed at
least in part to an incomplete convergence with respect to the number
of bands (e.g., 24 bands in Ref.~\onlinecite{Ku2002}) in combination
with the linearization error. In the case of Si we find that the latter
accounts for less than 0.03~eV, but we cannot rule out that it is
larger in other systems and must be taken into account in order to
obtain reliable $GW$ results.

Very recently van Schilfgaarde \emph{et al.}\cite{vanSchilfgaarde2005}
also reexamined the convergence of the self-energy and the quasiparticle
energies in Si, employing the same $GW$ algorithm\cite{Kotani2002}
as in this work. However, the eigenfunctions were generated by the
full-potential LMTO method with basis sets ranging from 50 to 185
orbitals, including additional MT orbitals located in the conduction-band
region. They obtained a $\Gamma_{25'v}$--$X_{1c}$ gap of 1.15~eV,
similar to our results in Table~\ref{cap:Quasip}; the slight discrepancy
is due to the different way of generating the basis functions used
to construct the self-energy. With the corrections for $\mathbf{k}$-point
convergence and screening due to the 2\emph{p} electrons, their best
estimate for the $\Gamma_{25'v}$--$X_{1c}$ gap, if the $GW$ approximation
is evaluated with LDA eigenstates, is 1.10~eV. Furthermore, Shishkin
and Kresse reported quasiparticle band gaps obtained with a new PAW
implementation in quantitative agreement with our results.\cite{Shishkin2006}

\section{Conclusions}

In this work we investigated the influence of the basis-set accuracy
in linearized all-electron methods on the $GW$ self-energy correction.
We showed that the addition of local orbitals defined as second and
higher energy derivatives of solutions of the radial Schr\"odinger (or
scalar-relativistic Dirac) equation constitutes an extension of the
LAPW basis set that allows a systematic improvement towards basis-set
completeness. In the case of silicon basis-set convergence of the
$GW$ results was essentially reached with the inclusion of second-derivative
local orbitals. It is the basis-set dependence of $\left\langle \Sigma^{\mathrm{c}}\right\rangle $
that is responsible for changes in the quasiparticle energies. As
all of them decrease towards basis-set completeness, the effect on
the relative transition energies is smaller than on the quasiparticle
energies themselves. The fundamental band gap increases by less than
0.03~eV. This makes the LAPW method a suitable quantitative reference
scheme for all-electron $GW$ calculations. A stronger basis-set dependence
in other systems or in the correction of higher bands is possible,
however. If convergence with respect to the number of bands is also
taken into account, the \textbf{$GW$} band gap turns out to be larger
and closer to pseudopotential values than previously reported all-electron
results, although a certain discrepancy still remains.

\begin{acknowledgments}
The authors acknowledge valuable discussions with Gustav Bihlmayer
and Mark van Schilfgaarde as well as financial support from the Deutsche
Forschungsgemeinschaft through the Priority Program 1145.
\end{acknowledgments}
\appendix*

\section{Scalar-relativistic equations\label{sec:app-scalrel}}

The radial scalar-relativistic Dirac equations for the large and small
components $p$ and $q$ of a free electron in a spherical potential
at energy $\epsilon$ are given by\begin{equation}
p'(\epsilon,r)=2M(\epsilon,r)q(\epsilon,r)+\frac{1}{r}p(\epsilon,r)\;,\end{equation}
\begin{equation}
q'(\epsilon,r)=-\frac{1}{r}q(\epsilon,r)+w(\epsilon,r)p(\epsilon,r)\end{equation}
with $M(\epsilon,r)=1+\left[\epsilon-V^{\mathrm{eff}}(r)\right]/\left(2c^{2}\right)$
and $w(\epsilon,r)=\left[l(l+1)\right]/\left[2M(\epsilon,r)r^{2}\right]+V^{\mathrm{eff}}(r)-\epsilon$,
where $V^{\mathrm{eff}}(r)$ is the spherical part of the Kohn-Sham
effective potential.\cite{Singhbook} Their $\nu$th energy derivatives
are given by\begin{eqnarray}
p^{(\nu)\prime}(\epsilon,r) & = & 2M(\epsilon,r)q^{(\nu)}(\epsilon,r)+\frac{1}{r}p^{(\nu)}(\epsilon,r)\nonumber \\
 &  & +\frac{\nu}{c^{2}}q^{(\nu-1)}(\epsilon,r)\;,\end{eqnarray}
\begin{eqnarray}
q^{(\nu)\prime}(\epsilon,r) & = & -\frac{1}{r}q^{(\nu)}(\epsilon,r)+w(\epsilon,r)p^{(\nu)}(\epsilon,r)\nonumber \\
 &  & +\nu w^{(1)}(\epsilon,r)p^{(\nu-1)}(\epsilon,r)\\
 &  & +\sum_{\mu=2}^{\nu}\left(\begin{array}{c}
\nu\\
\mu\end{array}\right)w^{(\mu)}(\epsilon,r)p^{(\nu-\mu)}(\epsilon,r)\;,\nonumber \end{eqnarray}
where $M^{(1)}(\epsilon,r)=1/(2c^{2})$ has been used. The $\nu$th
energy derivative $w^{(\nu)}(\epsilon,r)$ has the expression\begin{equation}
w^{(\nu)}(\epsilon,r)=(-1)^{\nu}\frac{\nu!l(l+1)}{2^{\nu+1}M(\epsilon,r)^{\nu+1}r^{2}c^{2\nu}}-\delta_{1\nu}\;.\end{equation}
In the nonrelativistic limit these formulas correspond to Eq.~(\ref{eqall:ulo-gen}).


\begin{thebibliography}{10}
\bibitem{Hedin1965}L. Hedin, Phys. Rev. \textbf{139}, A796 (1965). 
\bibitem{Hybertsen1985}M. S. Hybertsen and S. G. Louie, Phys. Rev. Lett. \textbf{55}, 1418
(1985). 
\bibitem{Godby1986}R. W. Godby, M. Schl\"{u}ter, and L. J. Sham, Phys. Rev. Lett. \textbf{56},
2415 (1986). 
\bibitem{Hybertsen1986}M. S. Hybertsen and S. G. Louie, Phys. Rev. B \textbf{34}, 5390 (1986). 
\bibitem{Godby1987}R. W. Godby, M. Schl\"{u}ter, and L. J. Sham, Phys. Rev. B \textbf{35},
4170 (1987). 
\bibitem{Aulbur2000}W. G. Aulbur, L. J\"onsson, and J. W. Wilkins, in \emph{Solid State
Physics}, edited by H. Ehrenreich and F. Spaepen (Academic, New York,
2000), Vol. 54, p. 1 and references therein.
\bibitem{Hamada1990}N. Hamada, M. Hwang, and A. J. Freeman, Phys. Rev. B \textbf{41},
3620 (1990).
\bibitem{Aryasetiawan1992}F. Aryasetiawan, Phys. Rev. B \textbf{46}, 13051 (1992)
\bibitem{Ku2002}W. Ku and A. G. Eguiluz, Phys. Rev. Lett. \textbf{89}, 126401 (2002). 
\bibitem{Usuda2002}M. Usuda, N. Hamada, T. Kotani, and M. van Schilfgaarde, Phys. Rev.
B \textbf{66}, 125101 (2002).
\bibitem{Kotani2002}T. Kotani and M. van Schilfgaarde, Solid State Commun. \textbf{121},
461 (2002).
\bibitem{Faleev2004}S. V. Faleev, M. van Schilfgaarde, and T. Kotani, Phys. Rev. Lett.
\textbf{93}, 126406 (2004). 
\bibitem{Arnaud2000}B. Arnaud and M. Alouani, Phys. Rev. B \textbf{62}, 4464 (2000).
\bibitem{Lebegue2003}S. Leb\`{e}gue, B. Arnaud, M. Alouani, and P. E. Bloechl, Phys. Rev.
B \textbf{67}, 155208 (2003). 
\bibitem{Ernst2005}A. Ernst, M. L\"uders, P. Bruno, W. M. Temmerman, and Z. Szotek (unpublished).
\bibitem{Tiago2004}M. L. Tiago, S. Ismail-Beigi, and S. G. Louie, Phys. Rev. B \textbf{69},
125212 (2004). 
\bibitem{Delaney2004}K. Delaney, P. Garc\'{\i}a-Gonz\'{a}lez, A. Rubio, P. Rinke, and
R. W. Godby, Phys. Rev. Lett. \textbf{93}, 249701 (2004); W. Ku and
A. G. Eguiluz, \emph{ibid.} \textbf{93}, 249702 (2004). 
\bibitem{Hohenberg1964}P. Hohenberg and W. Kohn, Phys. Rev. \textbf{136}, B864 (1964); W.
Kohn and L. J. Sham, \emph{ibid.} \textbf{140}, A1133 (1965).
\bibitem{Bross1990}H. Bross and G. M. Fehrenbach, Z. Phys. B \textbf{81}, 233 (1990).
\bibitem{Krasovskii1994}E. E. Krasovskii, A. N. Yaresko, and V. N. Antonov, J. Electron Spectrosc.
Relat. Phenom. \textbf{68}, 157 (1994).
\bibitem{Slater1937}J. C. Slater, Phys. Rev. \textbf{51}, 846 (1937); Adv. Quantum Chem.
\textbf{1}, 35 (1964).
\bibitem{Andersen1975}O. K. Andersen, Phys. Rev. B \textbf{12}, 3060 (1975); D. D. Koelling
and G. O. Arbman, J. Phys. F: Met. Phys. \textbf{5}, 2041 (1975).
\bibitem{Wimmer1981}E. Wimmer, H. Krakauer, M. Weinert, and A. J. Freeman, Phys. Rev.
B \textbf{24}, 864 (1981); M. Weinert, E. Wimmer, and A. J. Freeman,
\emph{ibid.} \textbf{26}, 4571 (1982).
\bibitem{Fleur}http://www.flapw.de
\bibitem{Hamann1989}D. R. Hamann, Phys. Rev. B \textbf{40}, 2980 (1989).
\bibitem{Krasowski1997}E. E. Krasovskii, Phys. Rev. B \textbf{56}, 12866 (1997). 
\bibitem{Singh1991}D. Singh, Phys. Rev. B \textbf{43}, 6388 (1991). 
\bibitem{vanSchilfgaarde2005}M. van Schilfgaarde, T. Kotani, and S. V. Faleev, cond-mat/0508295
(unpublished).
\bibitem{Shishkin2006}M. Shishkin and G. Kresse, Phys. Rev. B (to be published).
\bibitem{Singhbook}D. J. Singh, \emph{Planewaves, Pseudopotentials and the LAPW Method}
(Kluwer, Dordrecht, 1994).\end{thebibliography}
\end{document}